\documentclass[useAMS,usenatbib]{mn2e}
\usepackage{epsfig}
\usepackage{color}

\title[{The Colours of Satellite Galaxies}]{The Colours of Satellite Galaxies in Groups and Clusters}

\author[A.~S.~Font et~al.]{A. S. Font$^{1}$\thanks{E-mail:
andreea.font@durham.ac.uk},  R.~G.~Bower$^{1}$, I.~G.~McCarthy$^{1}$,
  A.~J.~Benson$^{2}$, C.~S.~Frenk$^{1}$, 
\newauthor J.~C.~Helly$^{1}$, C.~G.~Lacey$^{1}$, C.~M.~Baugh$^{1}$, S.~Cole$^{1}$\\
$^{1}$Department of Physics, University of Durham, South Road, Durham DH1 3LE\\
$^{2}$Theoretical Astrophysics, Caltech, MC130-33, 1200 E. California Blvd., Pasadena CA 91125, USA}
\begin{document}

\date{Accepted 2008, July 8. Received 2008, June 7; in original form 2008 March 19}

\pagerange{\pageref{firstpage}--\pageref{lastpage}} \pubyear{2008}

\maketitle

\label{firstpage}

\begin{abstract}
Current models of galaxy formation predict satellite galaxies in groups and
clusters that are redder than observed. We investigate the effect on the colours
of satellite galaxies produced by the ram pressure stripping of their hot
gaseous atmospheres as the satellites orbit within their parent halo. We
incorporate a model of the stripping process based on detailed hydrodynamic
simulations within the Durham semi-analytic model of galaxy formation. The
simulations show that the environment in groups and clusters is less
aggressive than previously assumed. The main uncertainty in the model is
the treatment of gas expelled by supernovae. With reasonable
assumptions for the stripping of this material, we find that satellite
galaxies are able to retain a significant fraction of their hot gas for several
Gigayears, thereby replenishing their reservoirs of cold, star forming gas and
remaining blue for a relatively long period of time. A bimodal distribution of
galaxy colours, similar to that observed in SDSS data, is established and the
colours of the satellite galaxies are in good agreement with the data. In
addition, our model naturally accounts for the observed dependence of 
satellite colours on environment, from small groups to high mass clusters.
\end{abstract}

\begin{keywords}
galaxies: clusters: general; galaxies: evolution -- galaxies: 
fundamental parameters: colours -- galaxies: luminosity function
\end{keywords}

\section{Introduction}

Recent multiwavelength imaging with the Sloan Digital Sky 
survey (SDSS; \citealt{york00}) has convincingly demonstrated 
that the colour-magnitude distribution of galaxies is bimodal  
\citep{strateva01,hogg02,blanton03,baldry06}. Other galaxy 
properties, such as the star formation rates, disk-to-bulge 
ratios, concentrations, stellar surface mass densities and gas 
mass fractions also show bimodal distributions 
\citep{kauffmann03,brinchmann04,balogh04,hogg04,kannappan04}. 
Elucidating the origin of these distributions requires 
understanding whether they arise in situ or whether they 
are environmentally driven (i.e., the ``nature'' or ``nurture'' 
dichotomy).
 
While they are not the only means by which theorists attempt to 
tackle the problem of galaxy formation, at present semi-analytic 
models offer the best hope for understanding the origin 
of the bimodality. The level of sophistication of these models 
has grown rapidly in recent years and so too has their ability 
to match a wide variety of observational data  
\citep{kauffmann93,lacey93,cole94,kauffmann99,somerville99,cole00,benson02,benson03,baugh05,croton06,bower06}.  The recent addition of feedback from active galactic 
nuclei (AGN) in the semi-analytic models (e.g., 
\citealt{granato04,bower06,cattaneo06,croton06,menci06,kang06}) 
has provided an explanation for three well-known, yet 
puzzling observational results: (1) the steep cut-off at 
the bright end of the  galaxy luminosity function; (2) the fact that 
most massive galaxies today tend to be dominated by old, red stellar 
populations; and (3) the absence of classical cooling flows in 
the centres of massive X-ray clusters.  In addition, some of 
these models (e.g., that of \citealt{bower06}, hereafter 
B06, which we adopt as the baseline model in this paper) are 
also able to reproduce the evolution of the K-band luminosity 
and the galaxy stellar mass functions out to high redshift (see 
\citealt{baugh06} and references therein for a more in-depth 
discussion of the successes and limitations of semi-analytic models).

In terms of galaxy colours, semi-analytic models with AGN 
feedback are able to produce a clear bimodal separation of 
colours at high luminosities, match the slope of the 
red and blue sequences and explain the absence of massive 
blue galaxies. However, it has recently become clear that these 
models are unable to match the relative numbers of 
faint satellite galaxies on the red and blue sequences. 
Specifically, \citet{weinmann06b} and \citet{baldry06} have 
shown that the faint satellite galaxies in groups and clusters 
modelled semi-analytically are on average too red in comparison to 
satellite galaxies of similar luminosity in the SDSS sample. 
Using DEEP2 data, \citet{coil08} have shown that a similar 
problem may exist at higher redshifts  ($z \sim 0.7-0.9$), where 
the semi-analytic models predict a stronger clustering of red 
galaxies than is observed. These results indicate that 
one or more physical processes (either internal or 
environmentally-driven) that affect the ability of the galaxies 
to form stars are not yet accurately accounted for in the 
models.  

In the present study, we concentrate on the effect of 
environmental processes on the colours of galaxies, particularly 
on the role that the ram pressure stripping of the hot gas from 
galaxy haloes plays.  At present, the semi-analytic models 
outlined above treat the stripping of the hot gas from the 
haloes of galaxies in a crude fashion, by assuming that the 
gaseous halo is completely and instantaneously stripped as the 
system crosses the virial radius of the (more massive) parent 
system and becomes a satellite.  
A direct consequence of this stripping is that the only fuel that 
is available for star formation once the galaxy becomes a 
satellite is that which resided in a cold disk when the galaxy 
first fell into the halo.  As a result, the satellite galaxy experiences a 
sharp decline in its star formation rate and its stellar 
population becomes red over time.  This process of cutting off 
the supply of hot gas that would otherwise cool and replenish 
the reservoir of cold, star-forming gas, is sometimes referred 
to as ``strangulation'' or ``starvation''  
(\citealt{larson80,balogh00}).  It is clear, however, that the 
maximally-efficient stripping assumed by most current semi-analytic 
models is not realistic, especially in cases where the mass of 
the satellite is comparable to that of the parent system \citep{wang07}. 

In a very recent study, \citet{kang08} investigate the effect of modelling the
stripping of satellites by an exponential decay. They adopt a stripping
factor that is independent of the properties and orbit of the parent and
satellite haloes and vary this to achieve the best match to the
observations. However, although this empirical approach is a step
forward, one expects the efficiency of the stripping to depend on the
orbit of the satellite and on the structural properties of both the
satellite and parent systems. In contrast, the model we present here aims 
to implement a physically motivated description for the stripping
effect, taking into account the relevant properties of the system.

Recently, \citet{mccarthy08} carried out a large suite of high 
resolution hydrodynamic simulations of the stripping of the 
hot gaseous haloes of galaxies and presented a simple, 
physically-motivated model that describes the 
simulation results remarkably well. These authors concluded 
that, typically, galaxies are able to retain a significant 
fraction of their hot haloes for long periods of time 
following virial crossing.  The results of these 
simulations are in qualitative agreement with a recent 
{\it Chandra} X-ray survey of massive 
satellite galaxies in hot clusters by Sun et al.\ (2007), who 
found that most massive satellites do indeed have detectable 
hot haloes (see also \citealt{jeltema08}).  The implication of 
these results is that the stripping of the hot gaseous haloes is 
much less efficient than previously assumed.  Consequently, some 
replenishment of the cold star-forming reservoir (via cooling of 
the hot halo) in the satellite galaxies is expected to take place, 
this prolonging star formation and resulting in bluer satellite galaxies.

In the present study, we incorporate the prescription for ram 
pressure stripping of hot haloes of \citet{mccarthy08} in the 
Durham semi-analytic code for galaxy formation, {\small GALFORM}, 
and show that this improvement in environmental physics brings 
the colours of satellite galaxies into much better agreement with the 
observational data.

The layout of the paper is as follows:  In Section  
\ref{sec:model} we present the details of the implementation of 
the new ram pressure model in {\small GALFORM}. In Section 
\ref{sec:results} we present results for the fraction of blue 
galaxies in the new model (\S \ref{sec:bluefrac}), discuss 
the environmental signatures in the colours of galaxies 
(\S \ref{sec:environ}) and present predictions for the red and blue
luminosity functions at redshift $z=0.1$ for galaxies of different type 
(\S \ref{sec:lumfunc}). In Section \S \ref{sec:discuss} we summarize 
and discuss our findings.

\section[]{A Semi-analytic Model of Galaxy Formation with Ram 
Pressure Stripping of Hot Gaseous Haloes}
\label{sec:model}

\subsection{The B06 version of {\small GALFORM}}
\label{sec:B06}

Apart from the minor changes described below, we use the 
version of the {\small GALFORM} semi-analytic code described in 
B06.  This version of {\small GALFORM} makes use of halo 
merger histories extracted from the {\it Millennium Simulation} 
with the techniques of \citet{helly03} (see also 
\citealt{harker06}). The {\it Millennium Simulation} 
\citep{springel_etal05} was carried out by the Virgo Consortium 
and it is one of the largest simulations of the growth of 
structure in the $\Lambda$CDM cosmology to date, containing 
approximately 10 billion dark matter particles in a cubic 
volume of $(500 h^{-1}$ Mpc)$^3$ with a particle mass of $8.6 
\times 10^{8} \, M_{\odot}/h$. Throughout the paper we adopt the 
same cosmological model assumed in the Millennium simulation (and 
in the B06 model) and quote our results in terms of the Hubble variable 
$h = H_0/100$~km$^{_1}$ Mpc$^{-1}$.

In terms of baryonic physics, 
the version of {\small GALFORM} developed by B06 has the same basic 
structure as in \citet{cole00}, but with the addition of an 
improved treatment of gas cooling and a new scheme for AGN feedback.  
In the present study, most of the basic parameters of the code 
(including, e.g., the efficiencies of supernovae and AGN feedback 
and the timescales for star formation and dynamical friction) are the 
same as those adopted in B06. However, in order to achieve better 
agreement with the zero-point colours of the observed red and blue 
sequences, we increase the value of the yield to $p=0.04$, which is 
a factor of two higher than the ``standard'' solar value adopted by
B06. With this change, the $^{0.1}(g-r)$ colours are redder by
$0.1$~mag compared with those in the B06 model (for galaxies on the 
red sequence). It also improves the metallicity of the intracluster
medium (ICM): whereas in the standard B06 model the metallicity of the
ICM was too low, $Z_{ICM} \simeq 0.15 Z_{\odot}$, in the model with 
double the yields $Z_{ICM} \simeq 0.3 Z_{\odot}$ (Bower et al. 2008), 
a value which is in better agreement with the X-ray measurements 
(e.g., \citealt{baum05}).  We note that large
yields such as these have been used in semi-analytical models in the
past in order to get better agreement between the model predictions
for galaxy colours and the observations
(see \citealt{kauffmann98,delucia04}; but see also the discussion of
\citealt{cole00} about increasing yields and consistency with 
stellar evolution models).  
 
One further minor change from the version of B06 concerns the  
division between haloes in the rapid cooling and hydrostatic regimes. 
B06 adopted a sharp transition between these two regimes.  In particular, 
the BO6 model assumes that radio-mode feedback is only effective in 
hydrostatic haloes with

\begin{equation}
t_{\rm cool}(r_{\rm cool}) > \alpha_{\rm cool}^{-1} t_{\rm 
ff}(r_{\rm cool}),
\label{eq:acool}
\end{equation}

\noindent where $t_{\rm cool}$ and $r_{\rm cool}$ are the cooling time 
and radius, $t_{\rm ff}$ is the free-fall time (as defined by
\citealt{cole00}) and $\alpha_{\rm cool}$ is an adjustable parameter 
(set in this model to $0.7$) that rescales the freefall time of the
 halo and has the effect of controlling the position of the break in the 
luminosity function\footnote{Equation (2) of B06 is incorrect;  
$\alpha_{\rm cool}$ should be replaced by $\alpha_{\rm 
cool}^{-1}$.}.  If the above condition is satisfied, it is then 
determined whether or not the central AGN is able to inject 
sufficient power to offset the energy being radiated away in 
the cooling flow and, if the available AGN power is greater 
than the cooling luminosity, the cooling flow is assumed to 
be completely quenched.

This strict dichotomy in halo properties has the undesirable
 effect that small changes in, for example, the gas mass
 fraction of haloes near the transition can greatly
 affect the star formation rate in their central galaxy. This
 leads to a small population of rapidly forming galaxies at
 the bright end of the blue sequence.  In the present paper,
 we have refined the criterion to make the transition more gradual.
 To do this, we reduce the cooling rate in haloes for which
 the cooling radius and effective freefall radius [$r_{\rm cool}(t)$
 and $r_{\rm ff,eff}(t) \equiv r_{\rm ff}(\alpha_{\rm cool} t)$] are close
 to each other. Specifically, if

\begin{equation}
 |r_{\rm cool}-r_{\rm ff,eff}| \le 0.5 \epsilon_{\rm cool}^{-1}r_{\rm
   cool},
\end{equation}

\noindent the net cooling rate is reduced:

\begin{equation} 
\dot{m}_{\rm cool,eff} = \dot{m}_{\rm cool} \left( 0.5 +
 \epsilon_{\rm cool} [1-r_{\rm ff}/r_{\rm cool}] \right), 
\end{equation}

\noindent where $\dot{m}_{cool}$ is the cooling rate in the absence of AGN
feedback. This function leaves the cooling rate unchanged if the
inequality above is not satisfied. For large values of $\epsilon_{\rm cool}$ 
the inequality applies for only a narrow range of 
$(r_{\rm cool}-r_{\rm ff,eff})$, and the behaviour of the B06 model is 
maintained. Here we adopt $\epsilon_{\rm cool}=10$, so that the
transition is still quite sharp, and the luminosity function is little 
affected.  The effect on the colours of galaxies at the bright tip of the blue
 sequence is, however, noticeable: by suppressing gas cooling in
 these objects, the tip of the blue sequence becomes redder,
 tending to curl up towards the red sequence.  This provides a
 slightly better match to observational data in this part of the
 colour-magnitude diagram (see \S \ref{sec:bluefrac}).

With these basic parameters, we run a model with a complete and 
instantaneous ram pressure stripping of the hot gaseous haloes 
of satellites (the ``default'' model) and a model where the ram 
pressure stripping of the hot haloes of satellites is calculated 
using the prescription of  \citet{mccarthy08} (henceforth 
called the ``hot ram pressure model'').  

\subsection{Implementation of Ram Pressure Stripping}
\label{sec:ram}

Below we give a brief description of the \citet{mccarthy08} 
ram pressure stripping model and how it is incorporated into 
{\small GALFORM}. The McCarthy et al.\ model is analogous to the 
original formulation of \citet{gunn72} for the stripping of a 
face-on cold disk, except that it applies to a 
spherical distribution of hot gas.  Specifically, the hot 
gaseous halo of the satellite will be stripped if the ram 
pressure ($P_{\rm ram}$) exceeds the satellite's gravitational 
restoring force per unit area ($P_{\rm grav}$):

\begin{equation}
\label{eq:ram}
P_{\rm ram} \equiv \rho_{\rm gas,p} v_{\rm sat}^2 > P_{\rm grav} 
\equiv \alpha_{\rm rp} \frac{G M_{\rm tot,sat}(r) \rho_{\rm 
gas,sat}(r)}{r} \, , 
\end{equation}

\noindent where $\rho_{\rm gas,p}$ is the gas density of the 
parent halo, $v_{\rm sat}$ is the velocity of the satellite with 
respect to this medium, $M_{\rm tot,sat}(r)$ is the total mass of 
the satellite within radius $r$ and $\rho_{\rm gas,sat}(r)$ is 
the density of the satellite's hot halo at this radius.  The
coefficient $\alpha_{\rm rp}$ is a geometric constant of order 
unity; McCarthy et al.\ find that $\alpha_{\rm rp} 
\approx 2$ gives the best fit to their hydrodynamic simulations.
Note that equation (\ref{eq:ram}) has not introduced any new free 
parameters into the {\small GALFORM} semi-analytical model, as 
$\alpha_{\rm rp}$ has been tuned to match the results of hydrodynamic 
simulations.

The satellite-centric radius where $P_{\rm ram} = P_{\rm grav}$ 
is referred to as the stripping radius and McCarthy et al.\  
assume that within this radius the satellite's gaseous halo 
remains intact while all gas exterior to this radius is 
stripped on approximately a sound crossing time.  This 
physically simple model has been shown to match the results of 
their high resolution hydrodynamic simulations\footnote{This 
includes both Lagrangian smooth particle hydrodynamic 
simulations with the GADGET-2 code \citep{springel05} and 
Eulerian adaptive mesh refinement simulations with the FLASH code 
\citep{fryxell00}.} remarkably well for a wide range of orbits, 
mass ratios, and structural properties.

In our implementation of the hot ram pressure model we fix the 
stripping radius by setting the ram pressure to its maximum 
value, which occurs at the pericentre of the satellite's orbit, 
and the gas is stripped at the instant it crosses the virial radius. 
We adopt this simplification as, by default, {\small GALFORM} does 
not track the full orbital evolution of the satellite galaxies 
(but see \citealt{benson03})\footnote{The Millennium simulation
  provides full dynamical information for subhaloes until they are 
tidally disrupted. In view of the limited resolution of the simulation, 
however, we choose to calculate the merging timescale of subhaloes
using the standard \citet{chandrasekhar43} formula rather than
following the subhalo orbits explicitly, as explained in BO6.
Since we are interested mainly in the broad statistical distribution of galaxy
colours, our results should not be affected by
the loss of information about individual orbits.}. In reality, the physical size of 
the stripping radius is a function of time, as the magnitude of 
the ram pressure varies along a non-circular orbit.  By setting 
the ram pressure to its maximum value, we overestimate the amount 
of stripping that occurs between the time when the satellite 
crosses the virial radius of the parent halo and when it reaches 
pericentre for the first time. However, for typical orbits this is not 
unreasonable, since this timescale is generally a small 
fraction of the total amount of time that the satellite spends 
in orbit about the parent halo.  In addition, since the full 
orbital evolution of the satellites is not followed, we neglect 
the effects of tidal heating and tidal stripping of the satellite 
system.  In terms of the removal of hot halo gas, however,
ram pressure stripping is a more efficient mechanism than tidal 
stripping for the vast majority of satellites (i.e., the 
stripping radius is typically smaller than the satellite's tidal 
radius; see \citealt{mccarthy08}).

The pericentres and the velocities at pericentre (both of which 
are required to compute the maximum ram pressure along the 
orbit) of the satellites are calculated by assuming the 
satellites have the same 2-D joint radial ($v_r$) and 
tangential ($v_\theta$) velocity distribution of infalling 
substructure as measured by \citet{benson05} from a large suite 
of Virgo Consortium cosmological simulations. \citet{benson05} finds that the following functional form describes the 2-D 
distribution well:

\begin{equation}
f(v_{r},v_{\theta})=a_{1} v_{\theta}
\exp{[-a_{2}(v_{\theta}-a_{9})^{2}-b_{1}(v_{\theta})(v_{r}-b_{2}(v_{\theta}))^{2}]}
\, ,
\end{equation}

\noindent where $a_{1}$, $a_{2}$, $a_{9}$, $b_{1}(v_{\theta})$ and
$b_{2}(v_{\theta})$ are coefficients and functions tabulated in
\citet{benson05}. We assume that this distribution is independent 
of halo mass and redshift. For each satellite we randomly sample this 
distribution, extracting a radial and tangential velocity 
pair which, in turn, allows us to calculate the energy and 
angular momentum (per unit mass) of the orbit.  We compute 
the pericentre radius and velocity of the satellite at 
pericentre by assuming that the orbital energy and angular 
momentum are conserved and by treating the satellite as a 
point mass orbiting within a Navarro-Frenk-White (NFW) 
\citep{navarro96,navarro97} potential with the same total 
mass and concentration as the parent halo (see, e.g.,
\citealt{binney87}).  By assuming that the orbital energy and angular 
momentum are conserved, we are ignoring the dynamical 
friction force acting on the satellites.  However, in cases 
where the dynamical friction force is strong, the satellite will 
quickly sink to the centre of the parent halo and, in any case, 
this process is important for the most massive satellites which are 
less affected by ram pressure stripping.

By default, the version of {\small GALFORM} presented in B06 
assumes that the gas density profiles of the hot haloes of all systems 
(including satellite systems) follow a $\beta$-model 
\citep{cavaliere76,cavaliere78}, with $\beta = 2/3$ (which 
provides a reasonable match to the X-ray surface brightness 
profiles of massive groups and clusters; Jones \& Forman 1984), a 
fixed core radius $r_c = 0.1 R_{\rm vir}$, and a normalization 
$\rho_0$ that is set to yield the correct mass of hot gas 
within the virial radius.  We adopt the same distribution in 
the present study.  We note, however, that we have also 
experimented with NFW density profiles for the hot gas 
(for both the satellite and parent systems), and find that 
the resulting fraction of satellite galaxies that are blue is 
quite similar to the case of the $\beta$-model. This is likely 
to stem from the fact that, typically, the hot gas is only stripped down 
to intermediate radii, where the $\beta$-model and the NFW profile 
are similar.

In addition to a hot tenuous atmosphere, each halo has a
reservoir of cold gas that developed from the cooling of
the hot atmosphere and which can potentially form stars.
As stars are formed, some of this cold gas is re-heated by feedback from,
for example, supernovae winds.  Prior to stripping, the re-heated gas 
is assumed to follow the same spatial and thermodynamic distribution as the hot 
gas in the halo (as in B06). We therefore treat the initial stripping
of re-heated gas in the same fashion as the hot gas, and transfer the
same fraction (as computed with the ram pressure stripping algorithm 
described above) from the re-heated gas of the satellite to the
re-heated gas reservoir of the parent halo. We record the total mass
stripped from the halo as $M_{\rm strip}$ and the fraction of the hot halo
that is stripped as $f_{\rm strip}$. Although this procedure is simple,
it is worth noting that in reality the spatial and thermodynamic
properties of the re-heated gas could be quite complicated (e.g.,
multiphase) and that it need not be distributed in the same way as the 
hot gas phase.  It would be worth revisiting this issue once a better 
treatment of the re-heated gas is incorporated into the semi-analytic 
model (Benson et al., in prep.).

The cooling rate of the remaining, unstripped gas
is calculated by cooling only the gas within the stripping radius and
assuming that stripping does not alter the mean density of gas within this
radius. We implement this by giving the satellite a nominal hot gas 
mass $M_{\rm hot}' = M_{\rm hot} + M_{strip}$ (where $M_{\rm hot}$ is the
true hot gas content of the halo) and applying the same cooling algorithm
as that used for central galaxies (except limiting the maximum cooling radius to
$r_{\rm strip}$ rather than $R_{\rm vir}$). This step ensures self-consistency in 
the treatment of the gas cooling between stripped and unstripped
galaxies, and therefore that the colours of satellites are predicted correctly.

   So far, we have only been considering the stripping of gas which
    was in the hot or re-heated phases at the time the satellite halo
    reached pericentre in the parent halo for the first
    time. However, as long as star formation continues in the
    satellite galaxy, supernova feedback will continue to re-heat gas
    and eject it from the satellite galaxy. The colours of satellite
    galaxies turn out to be sensitive to how much of this ``secondary'' 
    re-heated gas is stripped. The numerical simulations of
    \citep{mccarthy08} do not provide any direct information about
    the stripping of this re-heated gas, since they only treat the
    stripping of the initial hot halo. One can imagine
    several possible scenarios for the fate of the re-heated gas: 
    we adopt a picture
    in which the the re-heated gas is ballistically ejected as relatively 
    cold material that subsequently mixes or evaporates to become part 
    of the hot halo. In isolated galaxies, the re-heated gas is put 
    back with the same radial distribution as for the original
    hot gas halo. In satellite galaxies, we can consider 
    two extreme cases for the treatment of the secondary re-heated gas. 
    Continuing to apply the initial stripping criterion to the satellite
    galaxy as it orbits suggests that the re-heated gas should be
    stripped by the same factor $f_{\rm strip}$ as it is ejected from the
    galaxy. However, applying this at every
    timestep is too extreme: most of the re-heating occurs during the 
    outer part of the galaxy's orbit where the ram-pressure force 
    is small. It might therefore be more appropriate to consider a
    second case
    where little of this re-heated gas is stripped by ram-pressure
    effects. Further numerical simulations are required to elucidate
    which of these scenarios is physically more realistic and to
    determine a suitable parameterization of the time averaged stripping
    rate. At this point, the second (minimal stripping) case seems 
    more appropriate since the
    typical orbital timescale exceeds several Gyr, and is comparable
    to the timescale for the mass-growth of the parent halo 
    (see below).

    To allow for these uncertainties, we adopt a partially 
    empirical approach, and model the time-dependence of the hot gas
    mass in the satellite halo after its first pericentre passage as:

    \begin{equation}
    \dot{M}_{\rm hot} = (1 - \epsilon_{\rm strip}f_{\rm strip})
    \frac{M_{\rm reheat}}{\tau_{\rm reheat}} - \dot{M}_{\rm cool}
    \end{equation}

    Here ${M}_{\rm hot}$ is the mass of hot gas
    available for cooling and $\dot{M}_{\rm cool}$ is the rate at
    which it cools.  $M_{\rm reheat}$ is the mass of gas which has
    been re-heated by supernovae but not yet returned to the hot
    phase. In the absence of ram-pressure stripping, this re-heated
    gas is assumed to return to the hot phase on a timescale
    $\tau_{\rm reheat} = t_{\rm dyn}/\alpha_{\rm reheat}$, as in B06,
    where the value $\alpha_{\rm reheat}=1.26$ was chosen to match
    the observed galaxy luminosity function. The effect of
    ram-pressure stripping is described by the the term
    $\epsilon_{\rm strip}f_{\rm strip}$, where $f_{\rm strip}$ is the
    stripping factor already calculated for the initial pericentre
    using equation~(4), and
    $\epsilon_{\rm strip}$ is a new parameter (representing the time
    averaged stripping rate after the initial pericentre) which we
    adjust to fit
    the observations. The first (maximal) stripping case corresponds to
    $\epsilon_{\rm strip}=1$ if the orbital timescale is much
    shorter than the re-heating timescale, and the
    second case (minimal stripping) to $\epsilon_{\rm strip}=0$. The
    gas which is stripped from each satellite is added to the hot gas
    component of the parent halo.

    We find that if we take $\epsilon_{\rm strip}=1$, then the
    colour distribution of satellite galaxies looks very similar to
    the default model in which there is complete stripping of the
    initial hot gas and of the re-heated gas, with all satellites lying 
    on the red
    sequence. For $0.2 \la \epsilon_{\rm strip}<1$, the satellite
    colour distribution looks similar to the case $\epsilon_{\rm
    strip}=1$. For $\epsilon_{\rm strip} \la 0.2$, a blue sequence
    appears for the satellites, which results in better agreement
    with the observed colour-magnitude distribution. Values in the
    range $0 \leq \epsilon_{\rm strip} \la 0.2$ result in similar
    colour-magnitude distributions. For this paper, we adopt the
    value $\epsilon_{\rm strip} =0.1$.

The stripping of satellites is also affected by the growth of the halo in
which the satellite is orbiting. If we did not allow for the effect of
halo growth, a small satellite accreted at high redshift would not feel
the increasing ram-pressure effect as the parent halo grows in mass,
perhaps becoming a galaxy cluster by the present-day. In order to take
this effect into account,  we assign the satellite galaxy new orbital 
parameters and derive a new stripping factor every time the halo doubles 
in mass compared to the initial stripping event. If this factor exceeds
that applied previously, additional hot and re-heated material is
removed from the galaxy. This process tends additionally to
suppress on-going star formation in the satellites of massive haloes.

An additional physical effect that may be relevant but that is not 
accounted for in our model is that the (ram + thermal) pressure force 
exerted on the hot halo of the satellite could, in turn, raise the 
pressure force exerted on the cold gas and potentially stimulate 
additional star formation \citep{bekki03}. This would tend to result in bluer 
colours for the satellite galaxies.  The magnitude of this effect is 
presently unclear, however, and needs to be quantified 
either observationally or with the aid of hydrodynamic 
simulations that accurately include the effects of cooling, star 
formation, and feedback.

\section{Results}
\label{sec:results}

\begin{figure}
\includegraphics[width=9.cm]{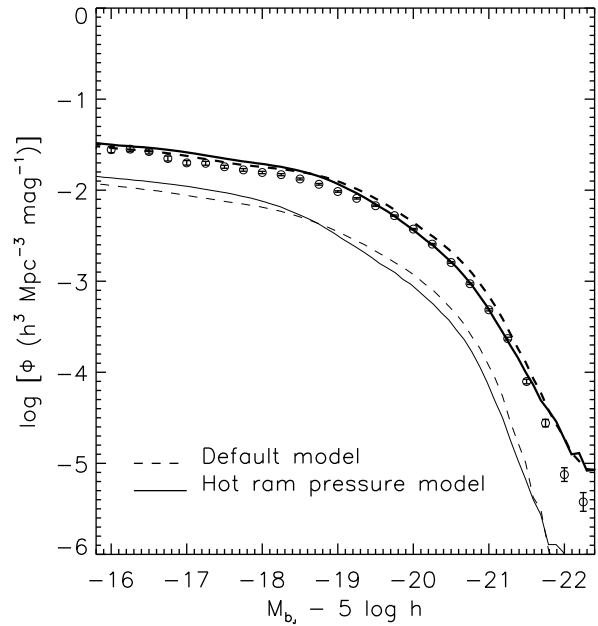}
\caption{\label{fig:lf}{Comparison of the predicted galaxy
luminosity functions of the default model with the hot ram
pressure stripping model at redshift $z=0$.  The data points represent 
observations from the 2dF Galaxy Redshift Survey \citep{norberg02}.
Magnitudes are $b_J$ (Vega) in both the models and the 
observations. Models use volume limited samples and the smallest 
resolved haloes have luminosities $\simeq 0.03 L_*$ (equivalent to 
$M_{b_J}-5\log h = -15.7$). Thick lines correspond to all galaxies and 
thin lines to satellite galaxies only.}}
\end{figure}

Since it is plausible that including a better treatment of the 
ram pressure stripping  may have a fairly significant impact on 
the observable properties of the galaxies, we first check if the 
predicted galaxy luminosity function is significantly altered.

In Fig.\ \ref{fig:lf} we compare the predicted galaxy luminosity 
function for the default\footnote{Our default model luminosity function is
essentially the same as that of B06, even after the minor changes 
outlined in \S \ref{sec:ram}} and hot ram pressure stripping models at
redshift $z=0$. The figure shows that our improved treatment of the 
ram pressure does not significantly alter the luminosity 
function.  This is primarily because ram pressure stripping 
preferentially affects satellite galaxies, whereas the total 
luminosity function (satellites + centrals) is dominated by 
central galaxies (see \S \ref{sec:lumfunc}).  In fact, even the 
luminosity function of satellite galaxies alone is only affected by 
a small amount (see thin lines in Fig. \ \ref{fig:lf}), however,  as we show 
below, the distribution of satellite colours at fixed luminosity 
is significantly altered.

\subsection{The blue fraction of galaxies}
\label{sec:bluefrac}

\begin{figure*}
\includegraphics[width=15.cm]{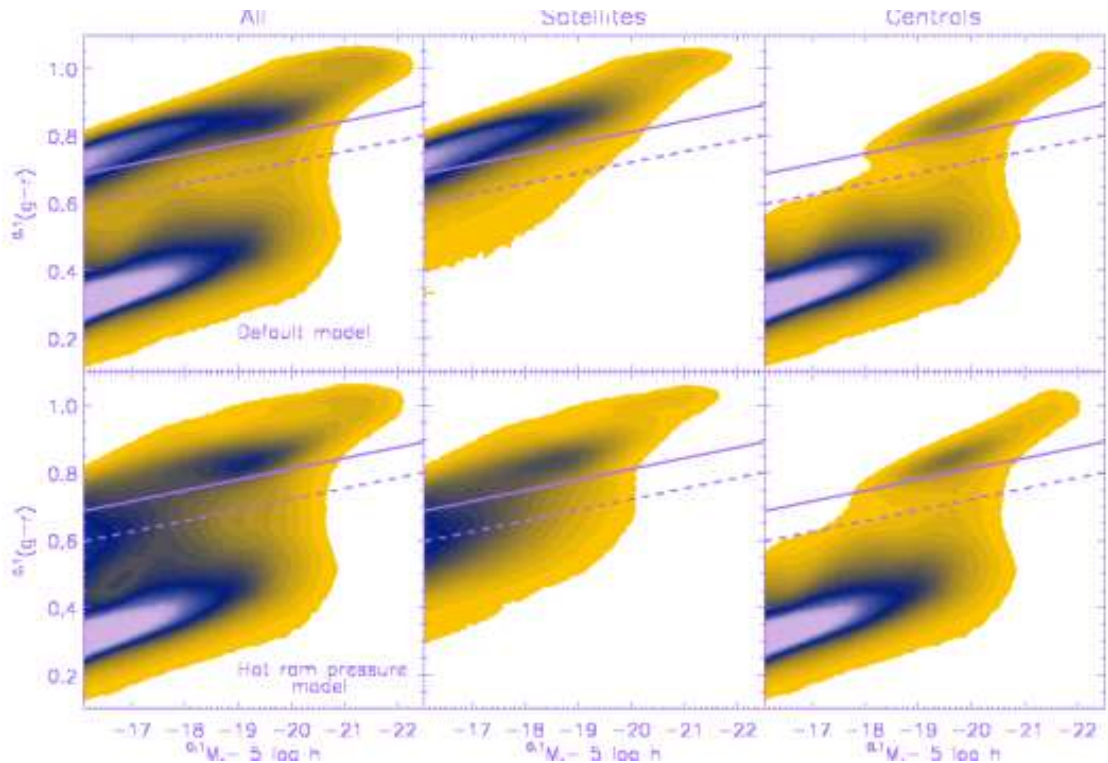}
\caption{\label{fig:cmd}{Colour-magnitude diagram (CMD) of 
$z =0.1$ galaxies in the default model (upper panels) and 
in the hot ram pressure stripping model (lower panels). The 
three panels in each row, from left to right, represent
the CMDs of all galaxies, satellites and
central galaxies, respectively. Magnitudes are SDSS (AB 
system) $^{0.1}g$ and $^{0.1}r$ at redshift $z=0.1$. The 
contours are spaced linearly in galaxy number density, starting 
from 500 per ($500 \ h^{-1}$ Mpc)$^3$ and increasing in levels 
of 500 per ($500 \ h^{-1}$ Mpc)$^3$. The solid line represents 
the colour cut adopted by \citet{weinmann06a} in order to separate blue and 
red galaxies in the SDSS. The dashed line represents an 
alternative colour cut (see text).}}
\end{figure*}

Fig.\ \ref{fig:cmd} shows the predicted colour-magnitude 
diagram (CMD) of present-day galaxies for the default model 
(upper panels) and for the hot ram pressure stripping model 
(lower panels).  The three panels in each row, from left to 
right, represent the CMDs of all galaxies, 
satellites only, and central galaxies only.  For a more 
convenient comparison with the SDSS data, the model predictions 
are output at redshift $z=0.1$ (which corresponds approximately  
to the median redshift of the SDSS sample) and the filters chosen
for analysis are $^{0.1}g$ and $^{0.1}r$ in the AB system. 

Fig.\ \ref{fig:cmd} shows, as expected, that changing the 
treatment of ram pressure stripping primarily affects the 
colours of satellite galaxies.  In particular, the 
distribution of $^{0.1}(g-r)$ colours at a fixed magnitude is 
broader for the hot ram pressure stripping model and a larger 
fraction of the satellites now have bluer colours.  The physical 
reason for this behaviour is simply that the retention of some 
of the hot gas in the halo of the satellite allows for the 
replenishment of the cold gas reservoir which, in turn, 
prolongs star formation in the satellite after its accretion 
onto the parent halo.  
Although the satellite galaxies eventually stop forming new
stars as they consume their gas reservoir, this process is much more 
protracted in the hot ram pressure model.  
As a result, the satellites cover a wider range of colours, filling 
in the region between the two main colour sequences.  The red sequence 
is also much less pronounced in the new model.

To quantify this change in the colours, we classify galaxies as 
being either `blue' or `red' by adopting the colour cut proposed 
by \citet{weinmann06a} for galaxies in the SDSS i.e.,

\begin{equation}
^{0.1}(g - r) = 0.7 - 0.032 \ (^{0.1}M_r - 5 \log h + 16.5) \, ,
\label{eq:colorcut}
\end{equation}

This cut, which is represented in the panels of Fig.\ 
\ref{fig:cmd} by solid lines, isolates reasonably well the red 
and the blue sequences in both the default and hot ram pressure 
models (see the top left and bottom left panels). 
This is similar to what is observed in the colour-magnitude
diagram of SDSS galaxies (see Figure 1 of \citealt{weinmann06a}; also Figure 1 
of \citealt{weinmann06b}).  For the SDSS data, the cut adopted by 
Weinmann et al. tends to isolate the red sequence at its base defined 
at the high luminosity end (as opposed to following the minimum between the red 
and blue sequences).  This effect is well reproduced by the solid line 
in Fig.\ \ref{fig:cmd}.

We note, however, that adopting exactly the same colour cut in our 
models as in the observations is not necessarily the best choice. 
A drawback of using the Weinmann et al. colour cut to define blue and
red fractions is that the results are quite sensitive to the precise position
of the red sequence. For this reason, we also employ a second cut 
(indicated by the dashed line in Fig.\ \ref{fig:cmd}), which intersects
the model's bimodal contours roughly at their minimum. In the
following, we will use by default the \citet{weinmann06a} cut and,
for illustrative purposes, show also results with the alternative cut.

\begin{figure*}
\centering
\includegraphics[width=18.cm]{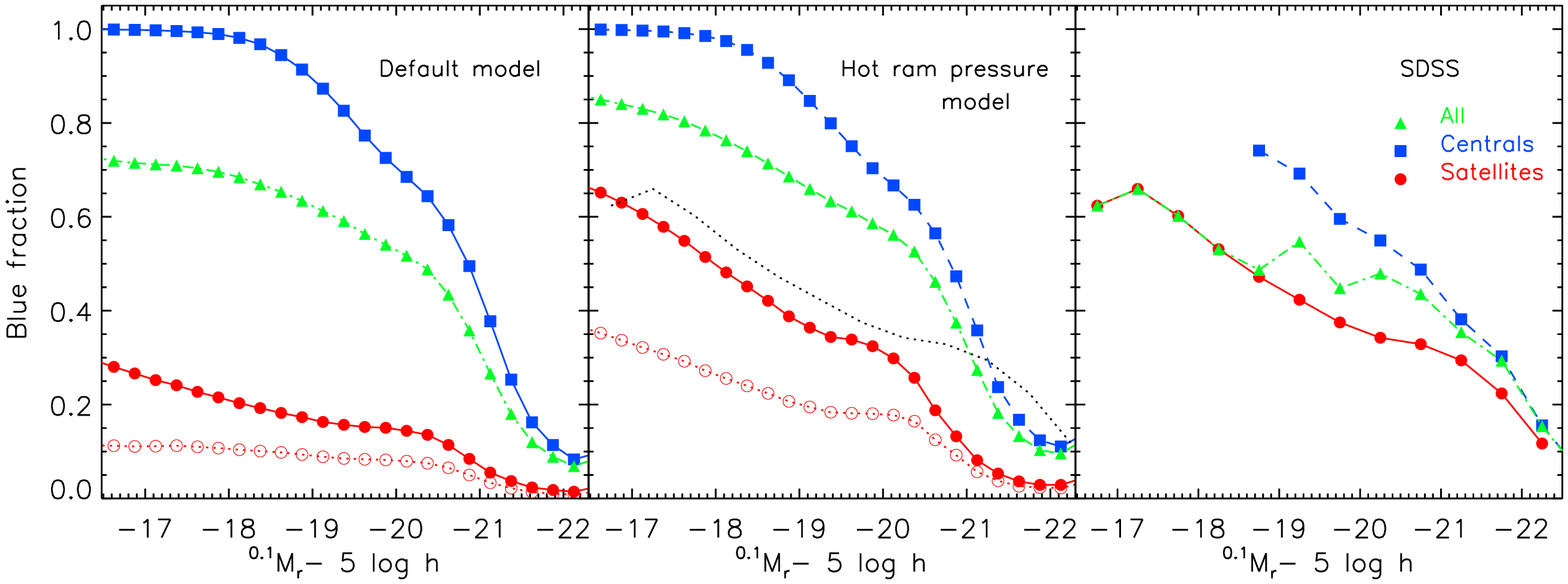}
\caption{\label{fig:fblue}{The fraction of blue galaxies per
$^{0.1}M_r$  magnitude. The left panel shows the 
default model, the middle panel shows the hot ram pressure 
stripping model. For comparison, we plot in the right panel 
the fraction of blue galaxies in the SDSS data as derived by 
\citet{weinmann06b}. The blue fractions are shown separately 
for satellites (red circles), central galaxies (blue squares) 
and all types (green triangles). The empty symbols 
represent the fraction of blue satellites with the 
alternative colour cut (see the dashed line in Fig.\ \ref{fig:cmd}). 
For better comparison, the black dotted line in the middle panel 
reproduces the blue fraction of SDSS satellites from the right panel.
All magnitudes are k-corrected to redshift $0.1$.}}
\end{figure*}

The left hand and middle panels of Fig.\ \ref{fig:fblue} show, 
for the default and hot ram pressure models respectively, the 
fraction of blue galaxies per $^{0.1}M_r$ magnitude for 
satellites (filled red circles), centrals (filled blue 
squares) and all types (filled green triangles) using 
the \citet{weinmann06a} cut.  Also shown is the fraction 
of blue satellites (open red circles) using the alternative 
colour cut.

Fig.\ \ref{fig:fblue} shows that the fraction of satellites that 
are blue in the hot ram pressure model is about twice as high as 
in the default model.  As discussed above, the actual fraction of blue 
galaxies depends on the colour cut adopted; however the relative 
difference in the blue fractions between the two models is largely 
independent of the cut. In particular, we find that using 
either the \citet{weinmann06a} or the alternative cut (or any 
in between these two), the fraction of satellite galaxies that are 
blue in the hot ram pressure model is approximately $2-2.5$ times 
larger than in the default model.

\begin{figure*}
\centering
\includegraphics[width=15.cm]{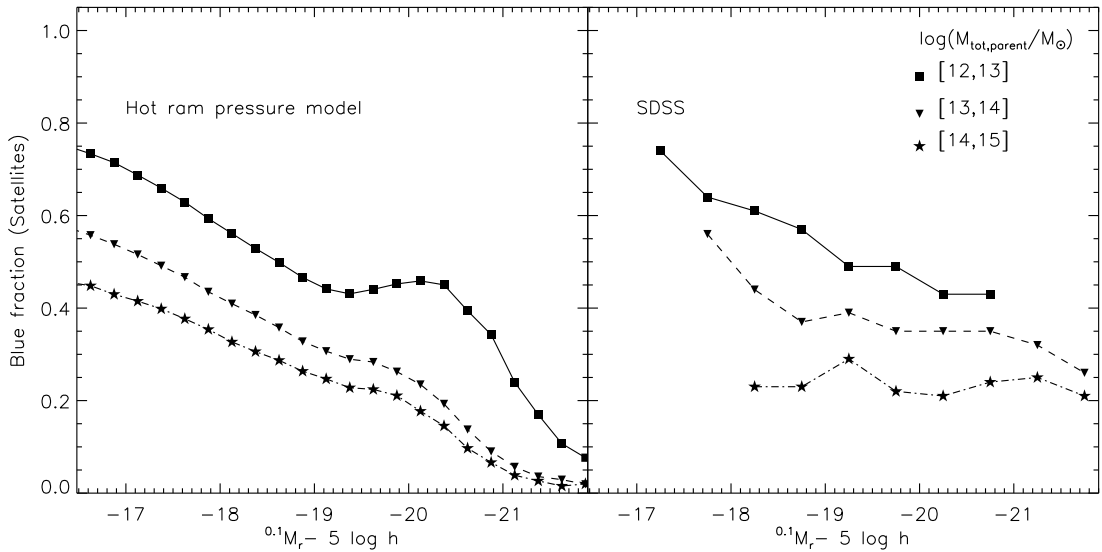}
\caption{\label{fig:fblue_lim}{The fraction of blue 
satellites per $^{0.1}M_r$ magnitude calculated using the Weinmann et al 
colour cut (eq. \ref{eq:colorcut}), in parent haloes of different mass.  
The hot ram pressure stripping model is shown in the 
left panel. Data from the SDSS group catalogue of \citet{weinmann06b} 
are shown in the right panel. In both panels magnitudes have been 
k-corrected out to redshift $0.1$.}}
\end{figure*}

How do the models fare in comparison to the observational data?  
In the right hand panel, we show the blue fraction of galaxies 
measured by \citet{weinmann06b} using an SDSS group catalogue 
(kindly provided by S. Weinmann). The fraction of satellite galaxies 
that are blue in the observations increases up to approximately $60\%$ 
at the faint magnitude end. Encouragingly, the newly implemented 
ram pressure model yields a  fraction of blue satellites that is in 
strikingly good agreement with the SDSS results.  The satellite 
blue fraction rises also to about $60$\% at the faint end and the 
overall trend in the blue fraction with luminosity is also 
close to that observed.  In contrast, the default model yields a 
fraction of blue satellite galaxies at most $\approx 30$\% 
(at faint magnitudes). This is similar to the fraction of blue 
satellites ($\sim 20$\%) reported for the \citet{croton06} 
semi-analytic model by \citet{weinmann06b} (see the bottom right 
hand panel of their figure 4). 

We also note that the fraction of bright central galaxies 
that are blue is in broad agreement with the observations. This 
is largely a consequence of including a prescription for 
AGN feedback into {\small GALFORM} which tends to halt star 
formation in the most massive galaxies.  Finally, we note that 
the semi-analytic model predicts many more faint central 
galaxies than observed in the SDSS group catalogue (independently of how the 
stripping of the hot gaseous haloes is treated). However, the lack of faint 
central galaxies is a selection effect of the group sample in the
SDSS: the source catalogue excludes groups for which no member is
brighter than $^{0.1}M_r= -19.5 +5\log h$, where $h=0.73$ 
\citep{weinmann06a}. Although we do not attempt to reproduce the 
mock groups catalogue in detail here, including this selection criterion 
tends to makes the satellite blue fraction shown in Fig. \ref{fig:fblue} 
greater by 0.1, further improving the match to the observations.

\subsection{Environmental dependence of colours}
\label{sec:environ}

Ram pressure stripping is an environmentally-driven process, 
hence one should expect variations in the fraction of blue 
satellites as a function of parent halo mass.  The left hand 
panel of Fig.\ \ref{fig:fblue_lim} shows the fraction of 
blue satellites in the hot ram pressure model per $^{0.1}M_r$ 
bin, calculated using the Weinmann et al. colour cut
(eq. \ref{eq:colorcut}), for different parent haloes mass 
ranges.  The lower masses, $10^{12} < M_{\rm tot,parent} < 10^{14} \, M_{\odot}$, 
correspond to group- type environments, whereas the larger 
masses,  $M_{\rm tot,parent}>10^{14} \, M_{\odot}$ correspond 
to massive clusters. For comparison, in the right 
hand panel we plot the blue fractions of satellites in the SDSS 
group catalogue of \citet{weinmann06b} split into the same 
parent halo mass bins.  

Reasonably good agreement with the data is obtained.  The new model
reproduces the overall trend that satellites of a given magnitude are
bluer if they reside in less massive systems.  The physical explanation
for this behaviour is due to the slighthy higher density of the hot gas in
more massive systems (because lower mass systems, on average, have
converted a larger fraction of their baryons into stars) and to the higher
orbital velocities of satellites in more massive host haloes.  The
latter effect is the dominant one, especially since the ram pressure
scales with the square of the satellite's velocity.  In particular,
if we ignore the weak dependence of hot gas density on halo mass, the
ram pressure scales simply as $P_{\rm ram} \propto v_{\rm sat}^2$.
Typically, $v_{\rm sat}$ is of order the virial circular velocity of the
parent halo, implying the ram pressure will scale roughly as $P_{\rm ram}
\propto M_{tot,p}^{2/3}$.  As a result, stripping is more extensive in
more massive parent systems, reducing the fuel supply for star formation
in satellites.

The model shows a slight increase in the fraction of 
blue satellite galaxies towards the bright magnitude end, i.e. the ``bump''
at $ -19 >^{0.1}M_{r}> -21$.  As Fig. \ \ref{fig:fblue_lim} shows, this 
effect is more pronounced for bright galaxies residing in low mass groups, 
$10^{12} < M_{\rm tot,parent} < 10^{13} \, M_{\odot}$, and likely reflects
a limitation of the way in which the transition between rapid cooling and
hydrostatic regimes is currently treated in the code (see \S \ref{sec:B06}).

The direct comparison between the halo masses in the
models and those inferred from the observations should be treated 
with some caution.  In the case of a theoretical model, the true mass
of the parent halo is known precisely, however in the case of the 
observational data one must make use of a mass proxy (e.g.,
\citealt{eke04}). In the particular case of the SDSS data, \citet{weinmann06b} 
use an empirical relationship between the total optical luminosity 
of a system and its mass.  The semi-analytic models, however, show 
considerable scatter in the relationship between these quantities.  
Ultimately, one would like to construct a mock survey from the theoretical 
models with similar characteristics to the SDSS catalogue, and
analyze the mock data the same way as the real data.  This is beyond 
the scope of the present paper and will be addressed in a future 
study.

We also note that similar results have been obtained from studies 
that attempt to distinguish low mass from high mass environments 
using other mass proxies. For example, \citet{hogg04} find that 
bluer galaxies in the local Universe typically reside in low galaxy 
density environments, whereas redder galaxies tend to live in high 
galaxy density environments.  This is fully consistent with the results of 
\citet{weinmann06b} and our own model predictions if one makes the 
reasonable assumption that the projected surface density of 
galaxies increases with increasing parent halo mass (e.g., 
\citealt{gladders07}).

\begin{figure*}
\includegraphics[width=12.cm]{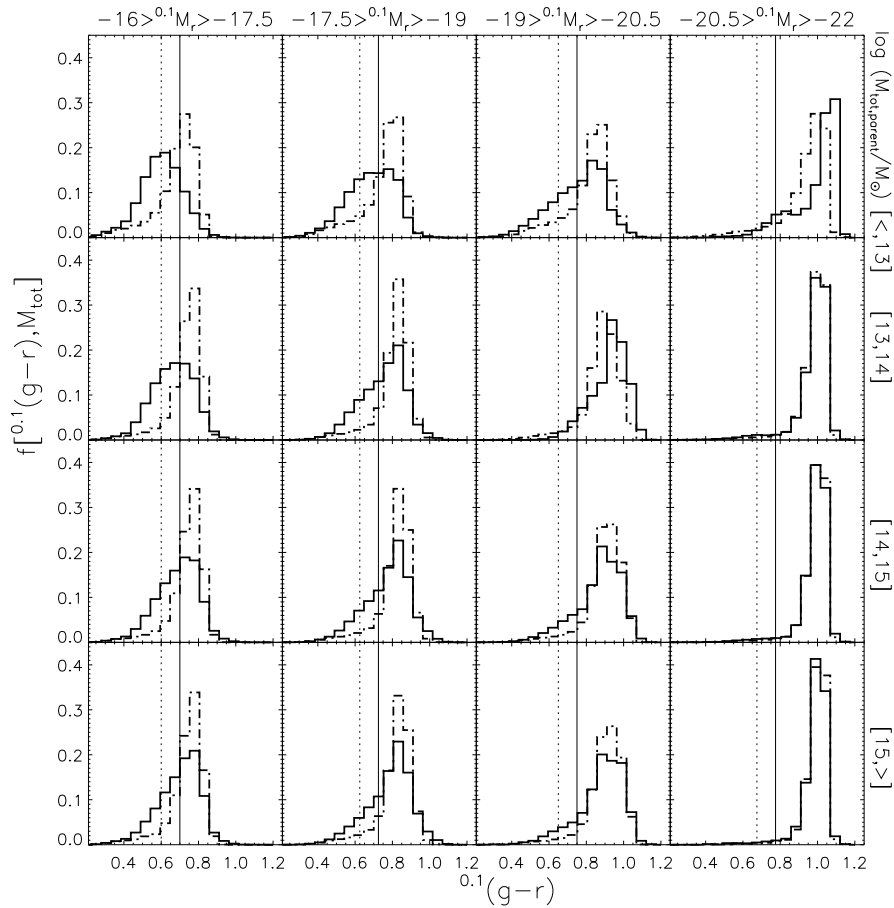}
\caption{\label{fig:hist}{Fraction of satellite galaxies 
($f[^{0.1}(g-r),M_{tot}]$) per $^{0.1}(g-r)$ colour bin, divided into 
different $^{0.1}M_r$ magnitude ranges and residing in parent haloes of 
different total mass. Results for the hot ram pressure
model are plotted with solid lines and for the default model with
dot-dashed lines. Magnitudes increase from left to right and parent halo 
mass increase from top towards bottom. Vertical lines correspond to the 
\citet{weinmann06a} colour cut (solid lines) and to the alternative
colour cut described in \S \ref{sec:bluefrac}(dotted lines).}}
\end{figure*}

Lastly, the colours are expected to depend not only on the mass 
of the parent halo but also on the intrinsic properties of the 
satellite.  To help disentangle these two factors, we plot in 
Fig.\ \ref{fig:hist}  colour  histograms (in $^{0.1}(g-r)$) for 
satellite galaxies in both the default (dashed lines) and 
the hot ram pressure (solid lines) models, divided into 
different $^{0.1}M_r$ magnitude and parent halo mass bins.  
In the new ram pressure model, satellites brighter than 
$^{0.1}M_r \sim -20$ tend to be red, independent of the parent halo
mass (i.e., environment). This is because internal processes, 
specifically AGN feedback, are the dominant mechanism for quenching the 
star formation in these systems. Meanwhile, as expected, 
low mass satellites in the same model (corresponding to magnitudes 
fainter than $^{0.1}M_r \sim -20$) tend to be bluer in low mass parent 
haloes and redder in massive clusters. The shapes of the histograms and
the overall trends are qualitatively similar to observational results
\citep{balogh04,weinmann06a} $-$ but note that the observations include
both satellites and centrals. For example, the highest luminosity galaxies 
are always dominated by the narrow red peak, regardless of environment,
a weak tail of blue galaxies only becoming visible in the lowest density
regions. Since we plot only the satellite galaxies here, a distinct blue
sequence is not seen; rather, the satellite galaxies increasingly occupy
transition colours at low halo masses where star formation
has been partially suppressed but not completely extinguished.
The results of our model suggest that, at the low halo mass end,
environmental processes are as important as the intrinsic physical 
processes in determining the colour of satellite galaxies.

\subsection{Luminosity functions of red and blue galaxies at $z=0.1$}
\label{sec:lumfunc}

\begin{figure*}
\centering
\includegraphics[width=18.cm]{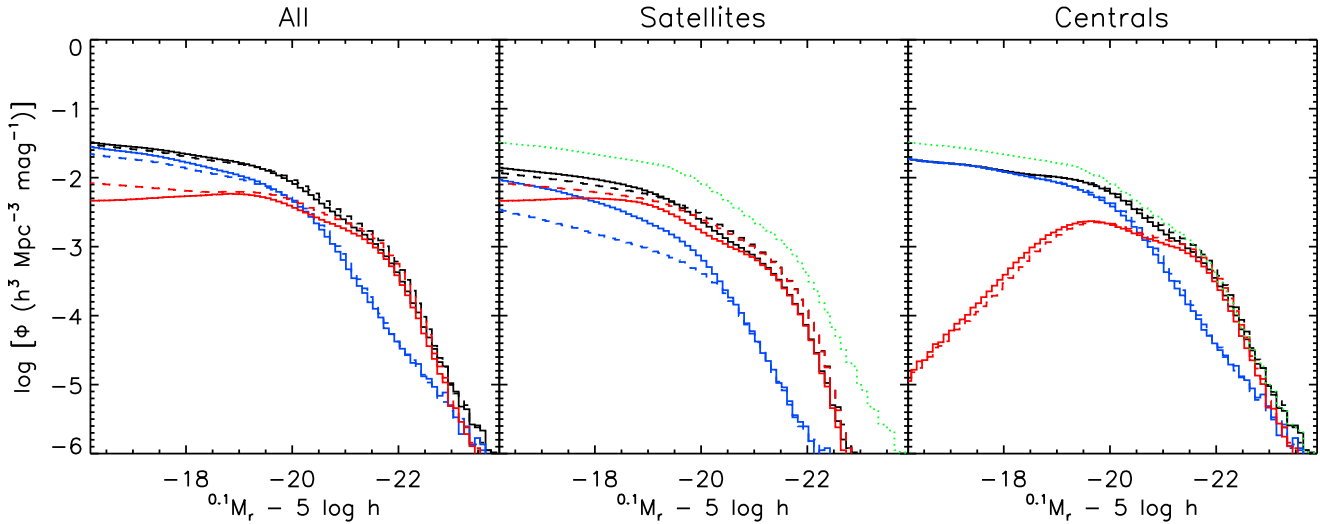}
\caption{\label{fig:lumfunc}{Luminosity functions in the $^{0.1}M_r$ band
    at redshift $z=0.1$. Panels include all galaxies (left), satellites
    (middle) and central galaxies (right). Solid lines correspond to the hot
    ram pressure model and dashed lines to the default model. 
    Black lines denote all galaxies in the (sub)sample, while red and 
    blue lines denote galaxies that are above and below the
    $^{0.1}(g-r)$ colour cut in equation \ref{eq:colorcut},
    respectively.  The green dotted lines in the middle and right hand 
   panels reproduce the total luminosity function of all galaxies for
   the hot ram pressure model (solid black line in left panel).
  }}
\label{fig:lumfunc}

\end{figure*}

As a starting point for future comparisons with SDSS and other data, we 
plot colour luminosity functions (LF) for the hot ram pressure model. 
Fig.~\ref{fig:lumfunc} shows the red and blue LFs in the $^{0.1}M_r$
band at redshift $z=0.1$, where colours are separated using the Weinmann et
al. cut.  Panels include all galaxies (left), satellites (middle) and 
central galaxies (right). Dashed lines represent the default model.
Current observational results for the luminosity functions are not
able to separate contributions of satellite and central galaxies
accurately over this range of luminosity due to the 
difficulty of robustly identifying faint central galaxies (However, see the
  recent HOD analysis of \citealt{brown08}). The comparison in 
Fig.~\ref{fig:lumfunc} highlights the origin of the variations of the 
blue fraction seen in Fig.~\ref{fig:fblue_lim}.

With the exception of the blue/red LF of satellites, the differences 
between the default and the hot ram pressure model are minor. Since the
centrals dominate in total number density, total LFs are similar for
both models (cf., discussion of Fig. \ref{fig:lf}).
When all galaxies are combined, the blue galaxy LF has a steeper 
faint end slope than the red counterpart, in agreement with observational
data (e.g., \citealt{baldry06}) \footnote{
    However, a study that appeared after our paper was submitted indicates that 
    the faint end slopes of the red and blue luminosity functions in
    the SDSS DR6 data are shallower than predicted by our model 
    \citep{montero08}. If confirmed, this may suggest that other
    physical processes unaccounted for in our model, 
    e.g. tidal stripping of stars \citep{henriques08}, may be
    responsible for the further flattening of the faint end slopes.}. By splitting the luminosity
function into central and satellite galaxies, we see that this is
driven by the rapidly increasing preponderance of blue central 
galaxies at faint magnitudes.  In contrast, the luminosity function of
central red galaxies drops by about three orders of magnitude between
$^{0.1}M_r$ of $-20$ and $-17$. This effect is driven by the AGN 
feedback in the model, and is independent of the stripping model.

The properties of satellites in the new model are much more dependent 
on the environmental physics included in the model. In the new model, 
the blue and red LFs reach similar values at the faint end 
(consistent with the results in Figure \ref{fig:fblue}
showing that at the faint end there are roughly similar numbers of red
and blue satellites).  The dramatic increase in the fraction of faint 
blue satellites can be seen by comparing the solid and dashed blue lines.
This is an alternative way of presenting the information in Fig.~\ref{fig:fblue_lim},
and underscores the importance of studying the environmental dependence of
galaxy properties in order to obtain a complete picture of galaxy formation
and evolution.

\section{Summary \& Discussion}
\label{sec:discuss}

Until now, most current semi-analytic models of galaxy 
formation have adopted a crude modelling of the ram pressure 
stripping of the hot gaseous halos of satellite galaxies.  In 
particular, they typically assume complete and instantaneous 
stripping of the hot gas halo when the galaxy first falls in, 
without regard for the galaxy's mass, orbit, or structural properties. 
This is at odds with results of hydrodynamic simulations, and also 
with recent X-ray observations of galaxies in massive clusters.  In 
the present study, we have improved the treatment of 
stripping by implementing the model of \citet{mccarthy08} 
(which has been shown to match simulations of ram pressure 
stripping to high accuracy) into the {\small GALFORM} 
semi-analytic model for galaxy formation. Although the initial
stripping event does not require us to add additional parameters 
to the model, subsequent stripping of the gas re-heated from the disk 
requires additional parameterization. We parameterize this process 
by assuming that most of the ejecta are retained in
the galaxy after the first pericentre passage.

We find that the newly implemented treatment of stripping leads to 
a significant improvement in the ability of the model to match 
the colours of satellite galaxies in the SDSS. The new model 
is also able to account for the environmental dependence of the 
colours of satellite galaxies.  Our results suggest that for the 
majority of satellite galaxies, this environmental 
process can be as important in modifying the galaxy colours as 
intrinsic processes, such as AGN or supernovae feedback, which 
operate within the satellites themselves. This finding is 
in broad agreement with previous studies that found that 
internal  processes that quench star formation do not seem 
capable of explaining the full range of colour and morphology 
data (e.g., \citealt{weinmann06a,baldry06}). 

Although we have only focused on the stripping of hot gaseous 
halos in the present study, other environmental processes may be 
relevant as well.  We now briefly review some of these and 
conclude that none of them appear to be as important as the ram pressure 
stripping of the hot haloes and the feedback already 
incorporated into our present model.

In cases where the ram pressure stripping of the hot halo is 
complete, some stripping of the cold gaseous disks may also 
occur \citep{gunn72,abadi99,quilis00}.  
However, \citet{okamoto03} and \citet{lanzoni05} have 
explored ram pressure stripping of disks in semi-analytic 
models and have concluded that the effect on the colours and star 
formation rates of satellite galaxies are minimal.  This most 
likely stems from the fact that disk stripping is only expected 
to be relevant for a minority of satellite galaxies whose 
orbits take them into the very centres of massive systems 
(e.g., \citealt{bruggen08}).  

Thermal evaporation of the hot gaseous haloes (and/or disks) 
could also be relevant \citep{cowie77}, but observational studies of 
bubbles and cold fronts in X-ray groups and clusters have placed strong 
constraints on the efficiency of conduction, concluding that it 
is strongly suppressed \citep{markevitch07,mcnamara07}.  
Turbulent stripping via the generation of Kelvin-Helmholtz and 
Rayleigh-Taylor instabilities at the interface between the hot 
gaseous halo of the satellite and the parent system is possible, 
but the timescale for this type of stripping is generally quite 
long (see \citealt{mccarthy08}).  

Other possibly relevant processes include viscous stripping 
(unfortunately at present the viscosity of the hot gas in groups 
and clusters is poorly constrained; \citealt{mcnamara07}), tidal 
effects on the gas as the result of the interaction with the gravitational 
potential of the parent halo \citep{byrd90,merritt83} or with 
the other satellites (i.e., mergers \citep{mihos95} 
or harassment \citep{moore96}), and shock heating of the 
satellite's hot gas as it falls in at transonic velocities.
However, \citet{mccarthy08} have argued that ram pressure 
stripping of the hot gaseous haloes is always more efficient 
than tidal stripping by the parent halo's potential or 
shock heating in cases where the satellite mass is less than 
about 10\% of the parent halo's mass. 

In this paper we have concentrated only on comparisons with data
at low redshifts ($z<0.1$). In a future study, we intend to 
test the model at other redshifts and compare the redshift evolution of galaxy 
colours with data from current deep surveys. Another important application of
the model is the study of the clustering of galaxies as a function of different 
physical quantities, such as colour. In addition to the large scale
dependence driven by the relative importance of AGN feedback and
ram pressure stripping as a function of halo mass, our model also predicts a 
radial dependence within a halo driven by the variation in the strength
of stripping with the galaxy orbit. The colour dependence of
small-scale clustering will constrain the model and perhaps suggest a
way to improve our treatment of the interaction between galaxies and their
environment.

\vskip 6pt
{\sf Galaxy catalogues for this model are available for download from this http URL: http://galaxy-catalogue.dur.ac.uk:8080/Millennium/}

\section*{Acknowledgments}

We thank Simone Weinmann for providing us the SDSS blue 
fraction data in electronic format. We are grateful to 
Simon White for
a careful reading of the manuscript and for useful suggestions. We also
acknowledge Michael Balogh, Michael Brown and David Wake for useful 
discussions. ASF is supported by a STFC Fellowship at the Institute for
Computational Cosmology in Durham. 
RGB acknowledges the support of a STFC senior fellowship.  
IGM acknowledges support from a postdoctoral 
fellowship from the Natural Sciences and Engineering Research Council 
(NSERC) of Canada. AJB acknowledges the support of the Gordon and Betty Moore 
Foundation.  This work was supported in part by a STFC rolling 
grant to Durham University.

\end{document}